# The Axion Dark Matter eXperiment

S.J. Asztalos, G. Caosi, C. Hagmann, D. Kinion, K. van Bibber
LLNL, Berkely, CA 94550
R. Bradley
NRAO, Charlottesville, VA 22903
John Clarke
UC & LBNL, Berkeley, CA 94720
C. Martin, P. Sikivie, N.S. Sulliva, D.B. Tanner
UF, Gainesville, FL 32611
M. Hotz, D. Lyapustin, L.J. Rosenberg, G. Rybka, A. Wagner
UW, Seattle, WA 98195

The Axion is a particle arising from the Peccei-Quinn solution to the strong CP problem. Peccei-Quinn symmetry breaking in the early universe could produce a large number of axions which would still be present today, making the axion a compelling dark matter candidate. The goal of the Axion Dark Matter eXperiment (ADMX) is to detect these relic axions through their conversion to photons in a strong magnetic field. Results are presented from a recent ADMX data-taking, along with plans for the next phase of ADMX, which will allow the experiment to explore a significant fraction of the favored dark matter axion mass and coupling phase space.

## 1. INTRODUCTION

Dark matter accounts for approximately 1/4 of the energy density of our universe, yet experiments have yet to find the elusive dark matter particle; such a discovery would obviously have tremendous physical implications. Dark matter is generally considered to be cold (non-relativistic) and non-baryonic, and within these constraints, the most promising candidates are weakly interacting massive particles and axions. Axions are particularly interesting because a laboratory experiment can be carried out with current technology that can detect or rule out axions at the expected level of abundance.

Axions arise from the Peccei-Quinn solution to the strong CP problem. As dark matter candidates, their mass and photon coupling are highly constrained by cosmological observations [1-6].

The Axion Dark Matter eXperiment (ADMX) searches for axion particles in the dark matter halo of our galaxy. These particles couple very weakly to photons, so a resonant cavity in a strong magnetic field is used to induce the axion-to-two-photons conversion. ADMX has excluded the 1.9 – 3.5 μeV axion mass region, and plans to cover the first full mass decade (to around 10μeV) after upgrades to the experiment are made.

## 2. PREVIOUS WORK

ADMX detects axions through their stimulated conversion to photons in a resonant high-Q cavity immersed in a strong magnetic field [7]. The cavity has rotatable rods that tune the resonant frequency of the cavity, and allow the experiment to scan over a range of approximately 500 Mhz to 1 Ghz in the TM010 resonant mode, which corresponds to an axion mass of about 1.9 – 3.5 μeV. Higher resonant modes allow scanning over higher mass ranges, but coupling to those modes is considerably weaker.

From the mid-1990's to the mid-2000's ADMX was able to rule out axions in this mass range, at couplings sensitive to dark matter models; to date, ADMX is the only experiment able to reach the model-sensitive region of mass and photon coupling phase space. Figure 1 shows the most recent published results from the experiment.

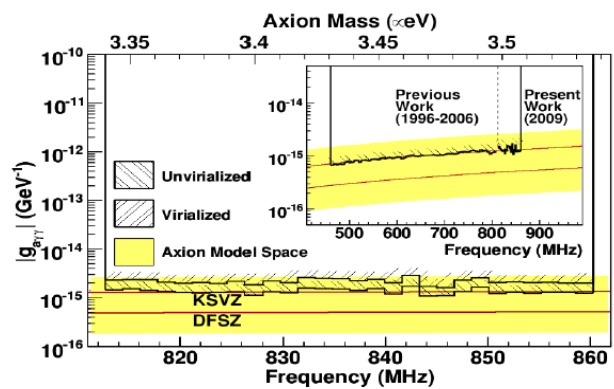

Figure 1: Axion couplings and masses excluded at the 90% confidence level by the experiment. The main figure shows the results from the Phase I upgrade and the inset has the limits from all data by ADMX.

## 3. CURRENT WORK - UPGRADES

ADMX recently underwent an upgrade that incorporated a superconducting quantum interference device (SQUID) in the receiver chain. The SQUID was employed as a high-gain, low noise amplifier, that could reach noise temperatures as low as 70 mK [8]. This temperature approaches the quantum limit, and is over an order of magnitude better than the 1.5 K HEMT amplifiers that were used in the original run of the experiment. ADMX will also include another detection antenna coupled to the $TM_{020}$ mode. Expected sensitivity is shown in figure 2, with the green and purple bands representing the two different channels.

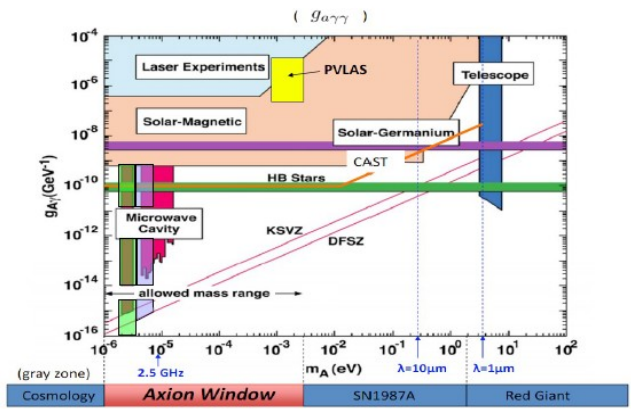

Figure 2: ADMX expected sensitivity to dark matter axions is shown in the green and purple bands. The green represents results expected from the $TM_{010}$ mode, and the purple from $TM_{020}$.

After the first decade of the experiment, the upgrades to ADMX were organized into two parts: ADMX Phase I would be a test of whether it would be possible to implement the SQUID in the middle of the ~ 7 T magnetic field, and Phase II would include a dilution refrigerator to get the SQUID down to mK temperatures.

Results from the search with a SQUID in the setup have been published [9], and further analyses from that data set are ongoing. In addition to placing the most stringent constraints on axion mass and coupling parameters, ADMX has been able to set limits on exotic axion-like particles [10,11].

## 4. CONCLUSION

Upgrades for the colder version of ADMX are currently underway. With the inclusion of a dilution refrigerator, physical temperature will drop to ~ 100 mK, and system noise temperature TNS is expected to be ~ 200 mK. Once fully upgraded, ADMX will be sensitive to, or be able to rule out, the first decade of axion masses (~ 1 - 10 μeV).

**Acknowledgments**

This work has been supported by the U.S. Department of Energy through grant numbers DE-FG02-97ER41029, DE-FG02-96ER40956, and DE-AC52-07NA27344 as well as through the LDRD program at Lawrence Livermore National Laboratory